\documentclass[preprint]{elsarticle}

\usepackage[colorlinks]{hyperref}
\usepackage{amsmath}
\usepackage{amssymb}
\usepackage{amsthm}
\usepackage[margin=3cm]{geometry}

\usepackage[font=small,labelfont=bf]{caption} 

\hypersetup{
    pdftitle = {Fast calculation of boundary crossing probabilities for Poisson processes},
    pdfauthor = {Amit Moscovich, Boaz Nadler},
    pdfkeywords = {Boundary crossing, Poisson process, Empirical process, Goodness of fit, Brownian motion, First passage}
}

\newcommand{\pr}[1]{\Pr\left[#1\right]}
\newcommand{\x}{{\bf x}}

\newcommand{\simiid}{\overset{i.i.d.}{\sim}}

\newtheorem{lemma}{Lemma}
\newdefinition{definition}{Definition}

\begin{document}

\title{Fast calculation of boundary crossing probabilities for Poisson processes}
\author[weizmann]{Amit ~Moscovich\corref{cor1}}
\ead{amit.moscovich@weizmann.ac.il}
\author[weizmann]{Boaz ~Nadler}
\ead{boaz.nadler@weizmann.ac.il}
\cortext[cor1]{Corresponding author}
\address[weizmann]{Department of Computer Science and Applied Mathematics, Weizmann Institute of Science, Rehovot, Israel.}

\begin{abstract}
    

The boundary crossing probability of a Poisson process with \(n\) jumps is a fundamental quantity with numerous applications. We present a fast $O(n^2 \log n)$ algorithm to calculate this probability
for arbitrary upper and lower boundaries. 
\end{abstract}

\maketitle

\section{Introduction}
Let $X_1, \ldots, X_n$ be $n$ i.i.d. random variables drawn from $U[0,1]$ and let
\( \hat{F}_n\) be their empirical cumulative distribution function,
\[
    \hat{F}_n(t) = \frac1n \sum_i \mathbf{1}(X_i \le t).
\]
Given two arbitrary functions $g,h:[0,1] \to \mathbb{R}$, we define the corresponding \emph{two-sided non-crossing probability} as
\begin{align} \label{eq:pr_binomial_no_cross}
    \pr{\forall t \in [0,1]: g(t) < \hat{F}_n(t) < h(t)}.
\end{align}
This probability plays a fundamental role in a wide range of applications, including the computation of $p$-values and power of sup-type continuous goodness-of-fit statistics
(\citet{Kolmogorov1933, Steck1971, NoeVandewiele1968, Noe1972, Durbin1973, KotelnikovaKhmaladze1983, FriedrichSchellhaas1998, KhmaladzeShinjikashvili2001});
construction of confidence bands for empirical distribution functions (\citet{Owen1995, Frey2008, Matthews2013});
change-point detection (\citet{Worsley1986});
and sequential testing (\citet{Dongchu1998}).
Note that many of these applications consider a more general case, where $X_1, \ldots, X_n \simiid F$ for some known continuous distribution $F$. However, this setting is easily reducible to the particular case $F=U[0,1]$
by  transforming the random variables $X_i \mapsto F(X_i)$ and the boundary functions as $g^*(t) = g(F^{-1}(t))$
and $h^*(t) = h(F^{-1}(t))$.

One popular approach is to estimate Eq. \eqref{eq:pr_binomial_no_cross} using Monte-Carlo methods.
In the simplest of these methods one repeatedly generates $X_1, \ldots, X_n \sim U[0,1]$
and counts the number of times that the inequalities $g(t) < \hat{F}_n(t) < h(t)$ are satisfied for all $t$.
This approach, however, can be extremely slow when the probability of interest is small and the sample size \(n\) is large. 

The focus of this paper is on the fast computation of the \emph{exact} two-sided crossing probability in Eq. \eqref{eq:pr_binomial_no_cross} given arbitrary boundary functions.
In the \emph{one-sided} case (where either $g(t) < 0$ or $h(t) > 1$ for all $0 \le t \le 1$),
Eq. \eqref{eq:pr_binomial_no_cross} can be computed in $O(n^2)$ operations
(\citet{NoeVandewiele1968, KotelnikovaKhmaladze1983, MoscovichNadlerSpiegelman2016}).
Even faster solutions exist for some specialized cases, such as a single linear boundary (\citet{Durbin1973}).
For general boundaries, however, essentially all existing methods require $O(n^3)$ operations (\citet{Steck1971, Durbin1971, Noe1972, FriedrichSchellhaas1998, KhmaladzeShinjikashvili2001})\footnote{The procedure of \citet{Steck1971} is based on the computation of an $n \times n$ matrix determinant and \citet{Durbin1971} is based on solving a system of linear equations.
Theoretically, using the Coppersmith-Winograd fast matrix multiplication algorithm, both methods yield
an $O(n^{2.373})$ solution. However this method is never used in practice because of the huge constant factors involved.}.

The main contribution of this paper is the introduction of a fast $O(n^2 \log n)$ algorithm
to compute the two-sided crossing probability for general boundary functions.
This is done by investigating a closely related problem involving a  Poisson process.
Specifically, let $\xi_n(t):[0,1] \to \{0, 1, 2\ldots\}$ be a homogeneous Poisson process of intensity $n$
and let $g,h:[0,1] \to \mathbb{R}$ be two arbitrary boundaries.
As noted in Section \ref{sec:binomial}, there is a well known reduction from the probability of interest in Eq. \eqref{eq:pr_binomial_no_cross}  to the following two-sided non-crossing probability, \begin{align} \label{eq:pr_conditional_poisson_no_cross}
    \pr{\forall t \in [0,1]: g(t) < \xi_n(t) < h(t)\ | \ \xi_n(1) = k}.
\end{align}
The key observation in this paper, described in Section \ref{sec:algorithm}, is that the recursive solution to Eq. \eqref{eq:pr_conditional_poisson_no_cross}
given by \citet{KhmaladzeShinjikashvili2001} can be described as a series of
at most $2n$ truncated linear convolutions involving vectors of length at most $n$.
Using the Fast Fourier Transform (FFT), each convolution can thus be computed in $O(n \log n)$ operations,
 yielding a total running time of $O(n^2 \log n)$.

In section \ref{sec:application} we present an application of the proposed method to the computation of $p$-values for a continuous goodness-of-fit statistic.
Comparing the run-times of our algorithm to those of \citet{KhmaladzeShinjikashvili2001} shows that our method yields significant speedups for large sample sizes.

Finally, since Brownian motion can be described as a limit of a Poisson process,
one may  apply our method to approximate the boundary crossing probability and first passage time of a Brownian motion, see for example \citet{KhmaladzeShinjikashvili2001}.
The latter quantity has multiple applications in finance and statistics (\citet{Siegmund1986, Bouchaud2013}).
In this case an accurate approximation may require a fine discretization of the continuous boundaries,
or equivalently a large value of $n$. Hence, fast algorithms are needed.
Furthermore, our approach can  be extended to higher dimensions,
where it may be used to quickly approximate various quantities related to Brownian motion in $2$ or $3$ dimensions.
\section{Boundary crossing probability for a Poisson process} \label{sec:algorithm}
\begin{figure}
    \centering
    \includegraphics[width=0.75\linewidth]{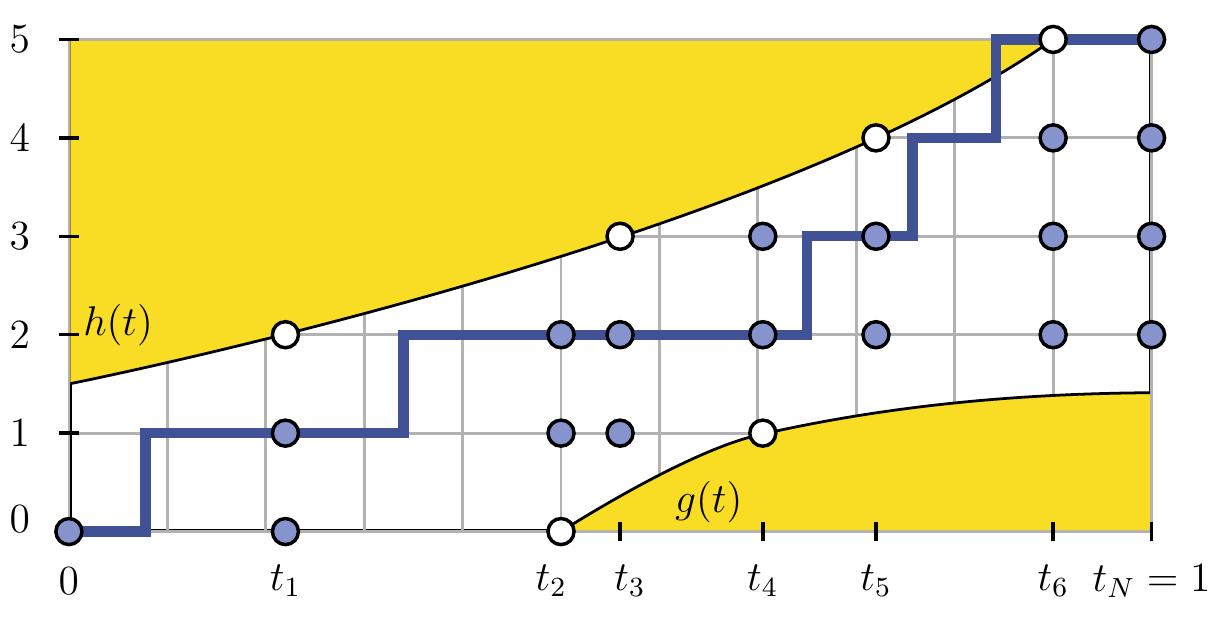}
    \caption{
        A two-sided boundary and a scaled empirical distribution $n\hat{F}_n(t)$ of $n=5$ samples.
        In this illustration $n\hat{F}_n$ happens to cross the upper boundary function $h(t)$.
        Empty circles mark the integer crossing points of $g(t), h(t)$ and determine
        discrete times $t_1 < \ldots < t_N=1$ which correspond to layers of a transition graph.
        Note that $n\hat{F}_n(t)$ crosses one of the boundaries if and only if it intersects an empty circle.
    }
    \label{fig:two_sided_crossing_illustration}
\end{figure}

In this section we describe our proposed algorithm for the fast computation of the
two-sided non-crossing probability of a Poisson process, given in Eq. \eqref{eq:pr_conditional_poisson_no_cross}.
We assume that $g(t) < h(t)$ for all $t \in [0,1]$ and that $g(0) < 0 < h(0)$, as otherwise the non-crossing probability
is simply zero.
Also, since the Poisson process is monotone, w.l.o.g. the two functions $g(t)$ and $h(t)$ may be assumed to be monotone non-decreasing.
We start by describing the recursion formula of \citet{KhmaladzeShinjikashvili2001}
whose direct application yields an $O(n^3)$ algorithm, and then show how to reduce the computational cost to $O(n^2 \log n)$
operations.

For every integer $i \in [0, g(1)]$, let $t_i^g = \inf\{t \in [0,1] : g(t) \ge i \}$
be the first time  the function $g(t)$ passes the integer $i$.
Similarly for every integer $i \in [h(0), h(1)]$, let $t_i^h = \sup\{t \in [0,1] : h(t) \le i\}$
be the last time the function $h(t)$ is bounded by  $i$.
Let $T(g) = \{t_i^g \}_{0 \le i \le g(1)}$ and $T(h) = \{t_i^h\}_{h(0) \le i \le h(1)}$ be the set of all integer crossing
times for the two functions.
As illustrated in Figure \ref{fig:two_sided_crossing_illustration},
a non-decreasing step function $f:[0,1] \to \{ 0, 1, 2, \ldots \}$
satisfies $ g(t) < f(t) < h(t)$ for all $t \in [0,1]$ if and only if it satisfies
these conditions at all discrete times $t \in T(g) \cup T(h) \cup \{1\}$.
Hence, to compute the probabilities in equations \eqref{eq:pr_binomial_no_cross}, and \eqref{eq:pr_conditional_poisson_no_cross},
it suffices to analyze these  inequalities only at a finite set of $N = |T(g) \cup T(h) \cup \{1\}|$ times.

\begin{definition} \label{def:Q}
    Let $\xi_n(t)$ denote a one-dimensional Poisson process with intensity $n$.
    For any $s \in [0,1]$ and $ m \in \{0, 1, 2, \ldots,n\}$, define $Q(s,m)$ as the probability that $\xi_n(s) = m$
    and that $\xi_n$ does not cross the boundaries $g(t), h(t)$ up to time $s$. i.e.
    \[
        Q(s,m) := \pr{\forall t \in [0,s]: g(t) < \xi_n(t) < h(t) \text{ and } \xi_n(s) = m}.
    \]
\end{definition}
Of particular interest are the values $Q(1,m)$ which correspond to Poisson processes that never cross the boundaries.
Clearly $Q(0,0) = 1$ and $\forall m > 0: Q(0,m) = 0$.
Let $t_1 < \ldots < t_N = 1$ denote the sorted set of times from $T(g) \cup T(h) \cup \{1\}$.
For any $i \in \{ 0, \ldots, N-1 \}$ and any $m \in \{ 0, 1, 2, \ldots \}$ the Chapman-Kolmogorov equations give
\begin{align} \label{eq:Qsm}
    Q(t_{i+1},m)
    =
    \begin{cases}
        \sum_{\ell} Q(t_i, \ell) \cdot\pr{Z_i = m-\ell} & \text{ if } g(t_{i+1}) < m < h(t_{i+1}) \\
        0 & \text{otherwise}
    \end{cases}
\end{align}
where $Z_i$ is a Poisson random variable with intensity $n(t_{i+1} - t_i)$ and the sum is taken over all $g(t_i) < \ell \le m$.
This formula was proposed by \citet{KhmaladzeShinjikashvili2001} in order to compute $Q(1,n)$.
All quantities up to the final time \(t_{N}=1\) can be computed recursively as follows:\ first calculate explicitly the probabilities $Q(t_1,0), \ldots, Q(t_1, n)$ at time $t_1$.
Next, calculate all probabilities at time $t_{i+1}$ using the quantities from time \(t_i\), and so on.
Since each $Q(t_i,m)$ is a sum of up to $m+1 \le n+1$ terms and since $N \le 2n+1$. the total run-time is at most $O(n^3),$
but may be smaller if the boundary functions $g(t), h(t)$ are close to each other.

Next, we describe a faster procedure.
Let $ Q_{t_i} = (Q(t_i,0), Q(t_i,1), \ldots, Q(t_i,n)) $
and let 
\(
    \pi_\lambda = (\pr{Z = 0},  \pr{Z = 1}, \ldots, \pr{Z = n}),
\)
where $Z \sim \text{Poisson}(\lambda)$.
The key observation is that the vector $Q_{t_{i+1}}$  in Eq. \eqref{eq:Qsm} is nothing but a \emph{truncated linear convolution} of the vectors $Q_{t_i}$
and $\pi_{n(t_{i+1}-t_i)}$. Hence  we may apply the  circular convolution theorem to compute it in the following fashion:
\begin{enumerate}
    \item Append $n$ zeros to the end of the two vectors $Q_{t_i}$ and $\pi_{n(t_{i+1}-t_i)}$,
    denoting the resulting vectors $Q^{2n}$ and $\pi^{2n}$ respectively.
    \item Compute the Fourier transform of the zero-extended vectors $\mathcal{F}\{Q^{2n}\}$ and $\mathcal{F}\{\pi^{2n}\}$.
    \item Use the convolution theorem to obtain the Fourier transform of the  convolution, 
    \[
        C^{2n} = \mathcal{F}\{Q^{2n} \star \pi^{2n}\} = \mathcal{F} \{ Q^{2n}\} \cdot \mathcal{F}\{\pi^{2n}\},
    \]
    where $\star$ denotes cyclic convolution and $\cdot$ denotes pointwise multiplication.
    \item Compute the inverse Fourier transform of $C^{2n}$ to yield the vector $Q_{t_{i+1}}$
    \begin{align*}
        Q_{t_{i+1}}(m) =
        \begin{cases}
            \mathcal{F}^{-1}\{C^{2n}\}(m) & \text{if }  g(t_{i+1}) < m < h(t_{i+1}) \\
            0 & \text{otherwise}.
        \end{cases}
    \end{align*}
\end{enumerate}
Using the FFT algorithm, each Fourier Transform takes $O(n \log n)$ time. Repeating these four steps for all times $t \in T(g) \cup T(h) \cup \{1\}$ yields a worst-case total run-time of $O(n^2 \log n)$.
However, it may be much lower if the functions $g(t)$ and $h(t)$ are close to each other.
For more details on the FFT and the computation of discrete convolutions, we refer the reader to \citet[Chapters 12, 13]{NumericalRecipes1992}.

\section{Boundary crossing probability for the empirical CDF} \label{sec:binomial}
We now return to the problem of calculating the probability in Eq. \eqref{eq:pr_binomial_no_cross},
that an empirical CDF will cross prescribed upper and lower boundaries. To simplify notation, we look at the scaled function  $n\hat{F}_n(t)$ instead of $\hat{F}_n(t)$, and similarly to the previous section, consider the probabilities
\[
    R(s,m) = \pr{\forall t \in [0,s]: g(t) < n\hat{F}_n(t) < h(t) \text{ and } n\hat{F}_n(t) = m}.
\]
Let $0 = t_0 < t_1 < \ldots < t_N = 1$ be as before,
and let
\[
    Z_{\ell,i} \sim \text{Binomial}\left(n-\ell, \tfrac{t_{i+1} - t_i}{1-t_i}\right).
    \]
The Chapman-Kolmogorov equations give the recursive relations of \citet{FriedrichSchellhaas1998}
\begin{align} \label{eq:Rsm}
    R(t_{i+1},m)
    =
    \begin{cases}
        \sum_{\ell} R(t_i, \ell) \cdot \pr{Z_{\ell, i} = m-\ell} & \text{if } g(t_{i+1}) < m < h(t_{i+1}) \\
        0 & \text{otherwise.}
    \end{cases}
\end{align}

In contrast to Eq. \eqref{eq:Qsm}, the expression for $R_{t_{i+1}}$, the vector of probabilities at time $t_{i+1}$,
is \emph{not} in the form of a straightforward convolution, and hence cannot be directly computed using the FFT.
While not the focus of our work, we note that by some algebraic manipulations,
it is possible to compute Eq. \eqref{eq:Rsm} using a convolution and an additional $O(n)$ operations.
Instead, we present a simpler construction that builds upon the results of the previous section.
To this end we recall a well-known reduction from the empirical CDF to the Poisson process (\citet[Chapter 8, Proposition 2.2]{ShorackWellner2009}):
\begin{lemma} \label{lemma:equiv}
    The distribution of the process $n \hat{F}_n(t)$ is identical to that of a Poisson process $\xi_n(t)$ with intensity $n$, 
    conditioned on $\xi_n(1) = n$.
\end{lemma}
According to this lemma, the non-crossing probability of an empirical CDF can be efficiently computed by
a  reduction to the Poisson case, since
\begin{align} \label{eq:binomial_noncrossing_problem}
    &\pr{\forall t: g(t) < n\hat{F}_n(t) < h(t)}
    =
    \pr  {\forall t: g(t) < \xi_n(t) < h(t) | \xi_n(1) = n} \\
    &= \frac{\pr{\forall t: g(t) < \xi_n(t) < h(t) \text{ and } \xi_n(1) = n}}{\pr{\text{Poisson}(n)=n}} 
    = \frac{Q(n,n)}{n^n e^{-n}/n!}\nonumber
\end{align}
and  $Q(n,n)$ can be computed efficiently, as described in Section \ref{sec:algorithm}.

\section{Computing p-values for goodness-of-fit statistics} \label{sec:application}
The results of the previous sections can be used to compute the $p$-value of
several two-sided continuous goodness-of-fit statistics such as Kolmogorov-Smirnov, and their power against specific alternatives.
Our algorithm may also be applied to one-sided statistics such as the Higher-Criticism statistic of \citet{DonohoJin2004}.

To this end, recall the setup in the classical continuous goodness-of-fit testing problem.
Let $x_1,x_2,\ldots,x_n$ be $n$ real-valued samples.
We wish to assess the validity of a null hypothesis that $x_1, \ldots, x_n$ are sampled i.i.d from a known
(and fully specified) continuous distribution function $F$ against an unknown and arbitrary alternative $G$,  
\begin{align*}
    \mathcal H_0: x_i \simiid F\quad vs. \quad \mathcal H_1: x_i \simiid G 
    \ \text{ with }\ G\neq F .
\end{align*}
Let $u_i = F(x_i)$ be the \emph{probability integral transform} of the $i$-th sample,
and  $u_{(1)} \le u_{(2)} \le \ldots \le u_{(n)}$ be the sorted sequence of transformed samples.
Under the null hypothesis, each $u_i$ is uniformly distributed in $[0,1]$ and therefore $u_{(i)}$
is the $i$-th order statistic of a uniform distribution.

A common approach to goodness-of-fit testing is to measure the distance of the different order statistics
from their expectation under the null.
A classical example is the Kolmogorov-Smirnov statistic 
$K_n :=\max\{K_n^-, K_n^+\}$, where $K_n^-$ and $K_n^+$ are the one-sided KS statistics, defined as
\[
    K_n^- = \sqrt{n} \max_{i=1,\ldots,n} \left( u_{(i)} - \frac{i-1}{n} \right),
    \quad
    K_n^+ = \sqrt{n} \max_{i=1,\ldots,n} \left( \frac{i}{n} - u_{(i)} \right).
\]
More generally,
given a sequence of monotone-increasing functions $r_1, \ldots, r_n:\mathbb{R} \to \mathbb{R}$
and a sequence of decreasing functions $s_1, \ldots, s_n:\mathbb{R} \to \mathbb{R}$,
one may define one-sided goodness-of-fit statistics by
\begin{align} \label{eq:def_R_S}
    R := \max_{i=1,\ldots,n} r_i(u_{(i)})
    \quad \text{and} \quad
    S := \max_{i=1,\ldots,n} s_i(u_{(i)})
\end{align}
and a two-sided statistic by
\begin{align} \label{eq:def_T}
    T := \max\{R, S\}.
\end{align}
Statistics of this form include
the supremum Anderson-Darling statistic and other weighted Kolmogorov-Smirnov statistics \cite{Kolmogorov1933, AndersonDarling1952},
the $R_n$ statistic of Berk and Jones \cite{BerkJones1979}
and Phi-divergence supremum statistics \cite{JagerWellner2007}.
Similarly, the one-sided Higher Criticism statistic of \citet{DonohoJin2004} and its variants follow the form of the one-sided statistic $S$ in Eq. \eqref{eq:def_R_S}.

It is easy to verify that $T < t$ if and only if $s_i^{-1}(t) < u_{(i)} < r_i^{-1}(t)$
holds for all $i$.
Therefore, the $p$-value of $T=t$  is equal to
\begin{align} \label{eq:pvalue}
    \pr{T > t | \mathcal{H}_0} = 1 -\pr{\forall 1 \le i \le n: s_i^{-1}(t) < U_{(i)} < r_i^{-1}(t)},
\end{align}
where $U_{(1)}, \ldots, U_{(n)}$ are the order statistics of $n$ draws from $U[0,1]$.
Define two functions \(g_{t}(x)\) and $h_{t}(x)$ as follows, 
\[
    g_{t}(x) = \sum_{i=1}^n \mathbf{1}(r_i^{-1}(t) \le x),
    \quad
    h_{t}(x) = \sum_{i=1}^n \mathbf{1}(s_i^{-1}(t) \le x),
\]
then the probability of Eq. \eqref{eq:pvalue} is equal to that of Eq. \eqref{eq:binomial_noncrossing_problem}
which we can compute in time $O(n^2 \log n)$.

\subsection{Simulation Results}

We evaluate the empirical run-time of our procedure
for computing $p$-values of the two-sided $M_n$ and one-sided $M_n^+$ goodness-of-fit statistics of \citet{BerkJones1979}.
These statistics have the form of equations \eqref{eq:def_R_S} and \eqref{eq:def_T} but with a minimum instead of a maximum
(see \cite[Section 3]{MoscovichNadlerSpiegelman2016}).

\begin{figure}
    \includegraphics[width=0.49\linewidth]{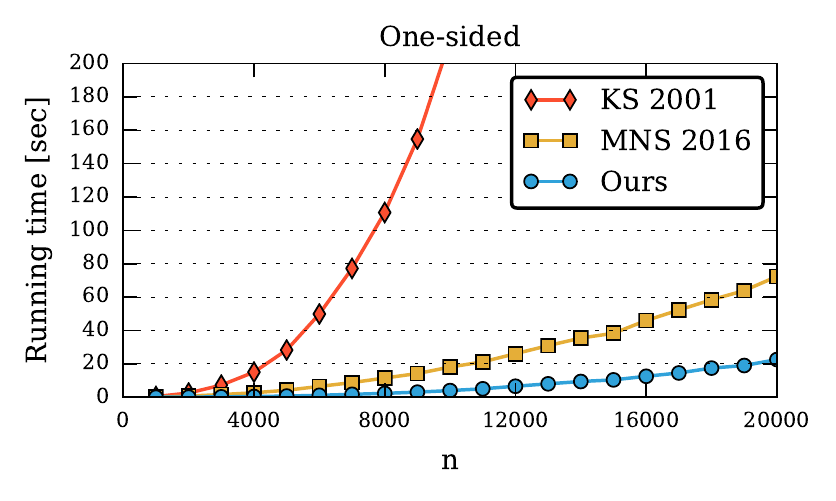}
    \includegraphics[width=0.49\linewidth]{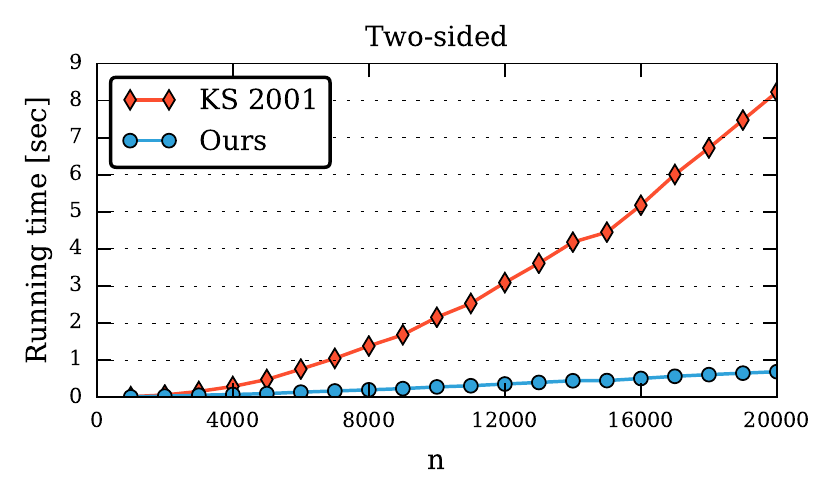}
    \caption{Runtime comparison of our algorithm compared to that of \citet{KhmaladzeShinjikashvili2001} (KS 2001)
    and to the one-sided method described in \citet{MoscovichNadlerSpiegelman2016} (MNS 2016).
    The boundaries were chosen such that the $p$-value of $M_n$, equal to its two-sided boundary crossing probability, is 5\%.
    Note that the one-sided case is much slower to compute.}
    \label{fig:running_times}
\end{figure}

To this end we wrote an efficient implementation of the proposed procedure
using the FFTW3 library by \citet{FrigoJohnson2005}
and compared it to a direct implementation of the \citet{KhmaladzeShinjikashvili2001} recursion relations (denoted "KS 2001").
In addition, we implemented the $O(n^2)$ \emph{one-sided} algorithm of \citet{MoscovichNadlerSpiegelman2016} (denoted "MNS 2016").
We find that both two-sided procedures are numerically stable using standard double-precision (64-bit) floating point numbers, even for sample sizes as large as $n=250,000$.
In contrast, the one-sided procedure \citep{MoscovichNadlerSpiegelman2016} requires a careful numerical implementation using extended-precision (80-bit) floating point numbers and breaks down completely for sample sizes $n>50,000$.
Figure \ref{fig:running_times} presents a runtime comparison
of the three algorithms
for computing one-sided and two-sided crossing probabilities\footnote{
C++ source code for all procedures is freely available at \url{http://www.wisdom.weizmann.ac.il/~amitmo}. The code was compiled using GCC 4.8.4 with maximum optimizations. The running times were measured on an Intel$^\circledR$ Xeon$^\circledR$  E5-4610 v2 2.30GHz CPU. 
}.

Somewhat counter-intuitively, the one-sided case is much more expensive than the two-sided case. This is made clear by examining Eq. \eqref{eq:Qsm} and noting that in the one-sided case the variable $m$ has a large valid range averaging around $n/2$,
whereas in the two-sided case this range is typically much smaller.
In all cases, our procedure is the fastest of all 3 methods.
Surprisingly, this is true even in the one-sided case where the $O(n^2)$ procedure of \citet{MoscovichNadlerSpiegelman2016} is asymptotically superior.

Finally, we note that for  large sample sizes, one may be inclined to forgo exact computation of $p$-values
and instead use the asymptotic null distribution of the particular test statistic in use (assuming it is known).
However, this does not always provide an adequate approximation, particularly as
in several cases the finite sample distribution converges very slowly to its limiting form.
Depending on the application, even the currently best known approximations may not be sufficiently accurate.
For more on this topic, see \citet{SiegmundLi2015}. 

\section{References}

\bibliographystyle{elsarticle-num-names}
\bibliography{crossing-probability}
\end{document}